\documentclass[10pt,conference]{IEEEtran}

\usepackage{geometry}
\usepackage{tikz}

\IEEEoverridecommandlockouts

\newgeometry{top=0.65in, bottom=0.9312in, left=0.52in, right=0.524in}

\usepackage{cite}
\usepackage{amsmath,amssymb,amsfonts}
\usepackage{graphicx}
\usepackage{textcomp}
\usepackage{xcolor}
\usepackage{algorithmicx}
\usepackage{algorithm}
\usepackage[bookmarks=false]{hyperref}
\usepackage{multirow}
\usepackage{subcaption}

\newcommand\copyrighttext{%
  \footnotesize This article has been accepted for presentation in \href{https://icmlcn2025.ieee-icmlcn.org/}{IEEE ICMLCN 2025}, but has not been fully edited. Content may change prior to final publication. \textcopyright 2025 IEEE. Personal use of this material is permitted. Permission from IEEE must be obtained for all other uses, in any current or future media, including reprinting/republishing this material for advertising or promotional purposes, creating new collective works, for resale or redistribution to servers or lists, or reuse of any copyrighted component of this work in other works.}
\newcommand\copyrightnotice{%
\begin{tikzpicture}[remember picture,overlay]
\node[anchor=north,yshift=-5pt] at (current page.north) {\fbox{\parbox{\dimexpr\textwidth-\fboxsep-\fboxrule\relax}{\copyrighttext}}};
\end{tikzpicture}%
}

\begin{document}

\title{SafeSlice: Enabling SLA-Compliant O-RAN Slicing via Safe Deep Reinforcement Learning\vspace{-0.3cm}}

\author{\IEEEauthorblockN{Ahmad M. Nagib\IEEEauthorrefmark{1}\IEEEauthorrefmark{3},
Hatem Abou-Zeid\IEEEauthorrefmark{2},
Hossam S. Hassanein\IEEEauthorrefmark{1}
}
\IEEEauthorblockA{\IEEEauthorrefmark{1}\textit{School of Computing}, 
\textit{Queen's University}, Canada, \{ahmad, hossam\}@cs.queensu.ca \\}
\IEEEauthorblockA{\IEEEauthorrefmark{2}\textit{Department of Electrical and Software Engineering}, 
\textit{University of Calgary},
 Canada, hatem.abouzeid@ucalgary.ca}
 \IEEEauthorblockA{\IEEEauthorrefmark{3}\textit{Faculty of Computers and Artificial Intelligence}, 
\textit{Cairo University},
 Egypt \\
 \thanks{This research was supported in part by the Natural Sciences and Engineering Research Council of Canada (NSERC) under Grant RGPIN-2021-04050.}
}

\\[-9.8ex]

}

\maketitle
\copyrightnotice

\begin{abstract}

Deep reinforcement learning (DRL)-based slicing policies have shown significant success in simulated environments but face challenges in physical systems such as open radio access networks (O-RANs) due to simulation-to-reality gaps. These policies often lack safety guarantees to ensure compliance with service level agreements (SLAs), such as the strict latency requirements of immersive applications. As a result, a deployed DRL slicing agent may make resource allocation (RA) decisions that degrade system performance, particularly in previously unseen scenarios. Real-world immersive applications require maintaining SLA constraints throughout deployment to prevent risky DRL exploration. In this paper, we propose SafeSlice to address both the cumulative (trajectory-wise) and instantaneous (state-wise) latency constraints of O-RAN slices. We incorporate the cumulative constraints by designing a sigmoid-based risk-sensitive reward function that reflects the slices' latency requirements. Moreover, we build a supervised learning cost model as part of a safety layer that projects the slicing agent's RA actions to the nearest safe actions, fulfilling instantaneous constraints. We conduct an exhaustive experiment that supports multiple services, including real virtual reality (VR) gaming traffic, to investigate the performance of SafeSlice under extreme and changing deployment conditions. SafeSlice achieves reductions of up to 83.23\% in average cumulative latency, 93.24\% in instantaneous latency violations, and 22.13\% in resource consumption compared to the baselines. The results also indicate SafeSlice's robustness to changing the threshold configurations of latency constraints, a vital deployment scenario that will be realized by the O-RAN paradigm to empower mobile network operators (MNOs).

\end{abstract}


\vspace{-1ex}
\section{Introduction}

6G networks will offer heterogeneous services that provide users with a full sensory experience by enabling the integration of the digital and physical worlds. This includes wearable devices, holographic communications and extended reality (XR) applications \cite{10054381}. To facilitate this, network slicing is used to partition the network into logically isolated slices that share underlying radio and computing resources \cite{9903386}. 

The open radio access network (O-RAN) architecture enables flexible slice management to meet diverse, sometimes conflicting, service level agreement (SLA) requirements, such as latency, throughput, and reliability \cite{o-ran-specification-slicing-architecture}. Its standardized open interfaces support data-driven network management, allowing mobile network operators (MNOs) to collect real-time data from the radio access network (RAN) \cite{9627832}. These interfaces also support the integration of intelligent slicing resource allocation (RA) controllers, utilizing machine learning (ML) capabilities. This facilitates optimizing resources dynamically based on network conditions and user demands.

Deep reinforcement learning (DRL)-based inter-slice RA algorithms have succeeded significantly in simulated environments. However, their direct application to physical systems, such as O-RANs, is often challenging \cite{9903386, ijcai2021p614}. In traditional DRL frameworks, the agent is primarily driven to explore any policy that optimizes resource consumption. In contrast, real-world immersive applications necessitate the additional requirement of satisfying certain constraints throughout the deployment process, such as the strict latency requirements of virtual reality (VR) applications.

Moreover, even if these constraints are considered in offline training, online learning will still be needed after deploying inter-slice RA agents in real-world O-RANs \cite{9903386}. This is typically required to allow DRL agents to tune their policies in real-time to adapt to the gaps between training and real-world environments. The system may undergo thousands of suboptimal learning steps in these situations before adapting. This potentially violates SLAs, especially at the instantaneous (state-wise) level, and degrades network performance. Furthermore, when an MNO modifies the defined SLA thresholds for a given slice, this can significantly affect the overall online learning performance. 

In this paper, we propose SafeSlice, a novel inter-slice RA approach that tackles both the cumulative and instantaneous latency constraints of slices while deployed in a live O-RAN setting. SafeSlice adapts to changes in network conditions and MNOs' configurations. The contribution of this research study can be summarized as follows:

\begin{itemize}
\item We design a multi-objective risk-sensitive DRL reward function that minimizes resource consumption while fulfilling the cumulative latency requirements of slices. We specifically propose a sigmoid-based function that includes parameters reflecting acceptable slices' SLAs. This enables penalizing the DRL-based slicing controllers for actions that get the system close to violating the defined SLAs.

\item We propose a safety layer to address the instantaneous latency constraints by projecting the RA actions to the nearest feasible actions. We particularly build a supervised ML model to predict the action cost in terms of latency. If an action is expected to violate the defined latency thresholds, the safety layer overrides it with the closest safe action expected to abide by the instantaneous latency constraints.

\item We advocate for safe DRL as a core component of intelligent DRL-based control of network functionalities in the O-RAN architecture. We propose training and deployment workflows in O-RAN's non-real-time (non-RT) and near-real-time (near-RT) RAN intelligent controllers (RICs), respectively. Such flows include collecting data to build the cost model and deploying SafeSlice in O-RANs.

\item We conduct an exhaustive performance study that supports multiple services, including real VR gaming traffic, to evaluate SafeSlice's performance against multiple defined baselines. Our approach not only adapts to changing and extreme traffic conditions but also to modifying latency threshold settings. The implementations of the proposed approach and baselines are publicly available\footnote{\href{https://github.com/ahmadnagib/safe-slice}{Available at https://github.com/ahmadnagib/safe-slice}} to facilitate research on trustworthy DRL in O-RAN.

\end{itemize}

To the best of our knowledge, this is the first study to 1) distinguish between and address instantaneous and cumulative constraints in DRL-based slicing, 2) accommodate changes in the acceptable SLA thresholds, and 3) evaluate the performance under varying and extreme network conditions, especially in an O-RAN deployment setting for immersive 6G applications.

The paper's remaining sections are structured as follows: In Section \ref{sec:related_safety}, we review the related work pertinent to our study. Section \ref{sec:proposal} provides a detailed description of the system model, and the proposed approach and O-RAN workflows. In Section \ref{sec:results}, the experimental setup, baselines and analysis of our performance study are presented. Finally, Section \ref{sec:conclusion} summarizes the key findings and presents potential future directions.

\section{Related Work}
\label{sec:related_safety} 

Safe reinforcement learning (RL) learns policies that maximize expected returns while ensuring acceptable system performance by adhering to safety constraints \cite{9903386}. A few studies have recently considered safety aspects when proposing a DRL-based approach to RA in network slicing. They primarily use a form of constrained RL. For instance, a variant of the actor-critic (AC)-based proximal policy optimization (PPO) algorithm is used in \cite{10.1145/3485983.3494850} to account for slice requirements during exploration. An artificial neural network (ANN) model is used to learn a cost function. If the model predicts performance degradation relative to slice requirements, it switches to a rule-based approach. However, such a rule-based approach is infeasible to implement in real-time as it performs a grid search. Moreover, the authors of \cite{10329913} propose a model-free multi-agent DRL approach that combines the AC and Lagrangian methods to reflect constraints in the objective function. Nevertheless, the constrained learning is applied during training only, which is insufficient to handle varying deployment conditions.

In \cite{10154267}, an algorithm that extends the soft actor-critic (SAC)-Lagrangian is used to optimize bandwidth allocation. The cost function is defined as the degradation in the quality of service (QoS). The authors incorporate the concept of domain randomization (DR) in training as a way to improve the generalization of the policy. Nonetheless, the study includes only one slice and treats different transmissions of one user as multiple users of the slice due to restrictions of the used tool. Similarly, the authors of \cite{9333595} propose a model-free constrained DRL algorithm based on SAC. Their main concern, however, is simultaneously handling both discrete and continuous action spaces for channel allocation and energy harvesting time division. As part of the learning policy update, they utilize a vector of Lagrangian multipliers to address the queue constraints that reflect latency and energy constraints. Nevertheless, the simulation is configured to use the same packet arrival rate for all slices with only a difference in packet size, which limits the practicality of the proposed approach.

More relevantly, the authors of \cite{9671840} propose a model-free RL-based slice management framework to maximize the throughput over time while satisfying latency and bandwidth constraints. They employ an adaptive constrained RL algorithm based on AC methods to deal with cumulative user dissatisfaction constraints. They also consider instantaneous latency and bandwidth constraints by adjusting the RA actions to align with the nearest feasible action. However, they do not include design, training, or deployment details of the approach to address the instantaneous latency constraints.

Despite the progress in adopting constrained DRL approaches in the context of slicing, the existing studies still experience limitations. Most reviewed studies do not distinguish between instantaneous and cumulative constraints. Moreover, such studies neither accommodate nor evaluate the performance of their approaches under changes in the acceptable SLA thresholds. Furthermore, the scenarios considered do not reflect the network being substantially overloaded or having significant changes in its conditions. Such situations are the primary triggers for the RL risky exploration challenge. Additionally, the live network deployment flows, especially in the context of O-RAN, have not been addressed in detail, even in O-RAN-oriented studies. The baselines in the reviewed studies either differ in their DRL design choices or lack sufficient detail in their descriptions. This results in unfair comparisons and bias towards the proposed approaches. Finally, the reviewed studies include some impractical assumptions and settings, such as the ones mentioned earlier.

\section{SafeSlice: Safe DRL-based Inter-Slice Resource Allocation for SLA-Compliant O-RAN Slicing}
\label{sec:proposal}

We propose SafeSlice, a DRL-based inter-slice RA approach that considers both cumulative and instantaneous latency constraints of the VR gaming slices. This is addressed by designing a risk-sensitive DRL reward function and building a supervised learning (SL) cost model that flexibly accommodates changes in the latency threshold settings, respectively. We also propose training and deployment workflows to enhance DRL safety in the O-RAN architecture.

\begin{figure*}[ht]
\centering
\includegraphics[width=0.8\linewidth]{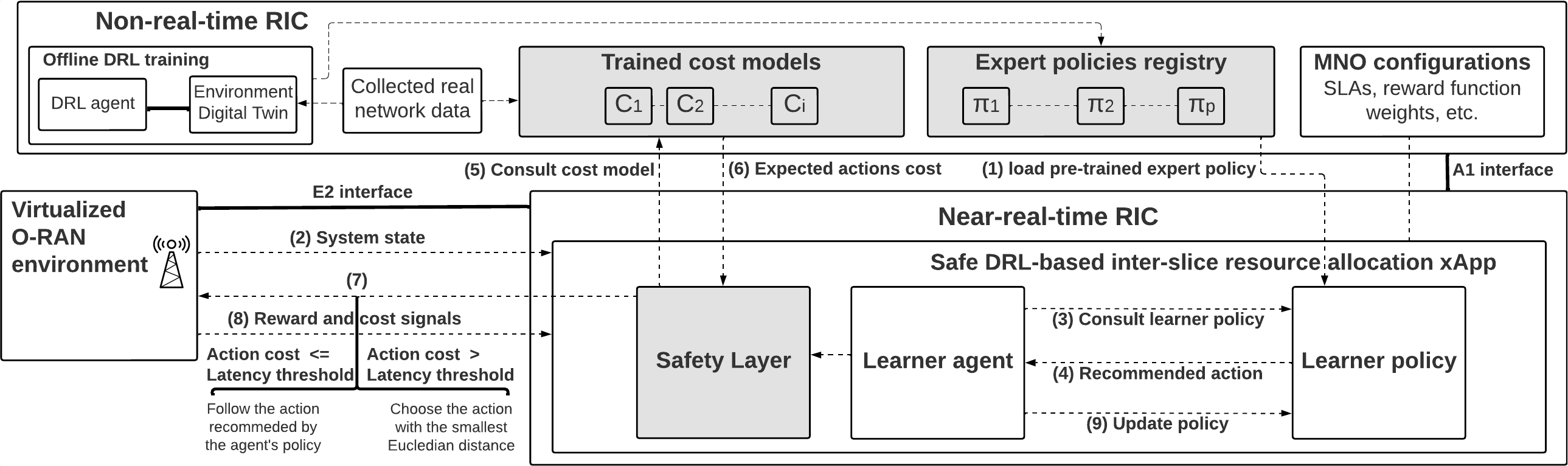}
\vspace{-1ex}
 \setlength{\belowcaptionskip}{-12pt} 
\caption{SafeSlice: The proposed safe DRL-based O-RAN slicing system.}
\label{fig:safe_system_diagram}
\end{figure*}

\subsection{System Model}
\label{model}

This paper addresses the downlink scenario within the radio access segment of network slicing. In O-RAN slicing, RA focuses on the efficient distribution of the limited physical resource blocks (PRBs) among the approved slices, ensuring that each slice's specific SLA requirements are met. We follow the model detailed below \cite{10437780}:

A base station (BS) facilitates a variety of services through a collection of virtual slices, denoted by $\mathcal{S}=\{1,2, \ldots, S\}$. These slices collectively utilize the available bandwidth, $B$. Each BS is connected to a set of user equipments (UEs), represented by $\mathcal{U}=\{1,2, \ldots, U\}$. For a given slice $s$, there are \( U_s \) associated users who generate a series of requests, $\mathcal{R}_s=\left\{1,2, \ldots, R_s\right\}$. The total demand, $D_s$, from these users is defined as:
\begin{equation}
D_s=\sum_{r_s \in \mathcal{R}_s} d_{r_s},
\end{equation}
where $d_{r_s}$ signifies the demand of a request, $r_s$, initiated by a user linked to slice $s$. Moreover, the contribution of a slice $s$ to the overall traffic demand of the BS at time $t$ is expressed as:
\begin{equation}
\label{contribution}
\kappa_s(t)=\frac{D_s(t)}{\sum_{i=1}^{\|S\|} D_i(t)}
\end{equation}

The allocation of PRBs among the admitted slices, $S$, is represented by the vector ${a \in \rm I\!R ^S}$, reflecting the percentage of bandwidth allocated to the available slices. At the beginning of a slicing window, \( \Omega_t \), a RAN slicing controller chooses a PRB allocation configuration, ${a}$, from the set of $A$ possible configurations, where $\mathcal{A}=\{1,2, \ldots, A\}$. This allocation influences the system's performance, which is measured by the actual resource consumption, $B_{\text{consumed}}\le B$, and is constrained by the latency requirements of the admitted slices. In our system, latency is predominantly affected by a queue maintained for each slice $s$ at a BS with a maximum capacity of \(q_{\text{max}}^s\).

Consequently, users of a given slice share the gNB resources allocated to such a slice equally. The number of packets arriving for a slice's users during the \(t\)-th transmission time interval (TTI) of a slicing window is denoted by \(H_{t}^s\), which is assumed to follow a distribution with an arrival rate and packet size of \(n_0\) bits that depend on the service supported by the slice. A slice's queue length at a TTI can be determined by \cite{9333595}:

\begin{equation}
q_{t}^s=\left(\min \left\{q_{t-1}^s+H_{t-1}^s-n_{t-1}^s, q_{\max }^s\right\}\right),
\end{equation}

where \(n_{t-1}^s\) represents the number of packets transmitted.

\subsection{Reinforcement Learning Mapping}
\label{sec:mapping}

The state of the system is represented by each slice's contribution to the total BS traffic during the preceding slicing window, specifically:

\vspace{-1.25ex}
\begin{equation}
\label{eq:state-represenation}
\kappa = (\kappa_{1}, \ldots, \kappa_s, \ldots, \kappa_{S}) 
\end{equation}
\vspace{-3ex}

At the beginning of each slicing window, the DRL agent perceives this state and subsequently determines the PRB allocation for every slice by selecting an appropriate action, defined as:

\vspace{-1.25ex}
\begin{equation}
\label{eq:action}
a = (b_{1}, \ldots, b_{s}, \ldots, b_{S}), 
\text{ subject to } b_{1} + \ldots + b_{S} \le B 
\end{equation}

We model the problem of inter-slice RA as a constrained Markov decision process (CMDP) to minimize resource consumption while considering the instantaneous and cumulative latency constraints for the VR gaming service, as explained in the next section. The cumulative, or trajectory-wise, constraint strives to guarantee that the average of cost when following slicing policy $\tilde{\pi}$ is below the latency threshold $\epsilon$ over a long sequence of states and actions during which an MNO is enforcing $\epsilon$. On the other hand, the instantaneous, or state-wise, constraint seeks to keep the cost associated with each network state along the slicing policy trajectory below the latency threshold $\omega$ \cite{ijcai2021p614}.

\subsection{Proposed Approach}
\label{sec:flow}

In this study, unlike most of the slicing RA literature, we simultaneously address the cumulative and instantaneous latency constraints. The main components of SafeSlice are highlighted in grey in Fig. \ref{fig:safe_system_diagram}. Section \ref{sec:training_workflow} and Section \ref{sec:safety_deployment_flow} detail the training and deployment workflows of our approach and its interaction with the different elements of the O-RAN environment.

\vspace{2pt}

\subsubsection{Risk-Sensitive Multi-Objective Reward Function}
 \label{sec:multi-objective_reward-function}

We design a multi-objective reward function that minimizes resource consumption while fulfilling the cumulative constraints. To address the cumulative latency constraints, we embed them in the reward function as a sigmoid-based component that includes parameters to reflect the acceptable SLAs for each slice. This mechanism imposes penalties on the DRL agent for actions that drive the system towards breaching the predefined acceptable latency limits of each slice, as illustrated below:

\begin{equation}
\label{eqn:safetyrewardequation}
\begin{aligned}
\scriptstyle
R \,= \biggr[\ \!\!W_{u}\ \!\times\ \!\Bigr(1 - \frac{\mathrm{\sum\limits_{s=1}^{\|S\|} b_{s}} }{\mathrm{B}}\Bigr) \!\biggr] + \biggr[\ \!\!W_{l}\ \!\times\ \!\sum\limits_{s=1}^{\|S\|} W_{s}\ \!\times\ \!\frac{\mathrm{1} }{\mathrm{1} + e^{\ c_{1,s}\ \times\ (\ {l}_{s}\ -\ c_{2,s}\ )} } \biggr]
\end{aligned}
\end{equation}

The weights, $W_{u}$ and $W_{l}$ $\in [0,1]$, reflect the importance of the goals of minimizing resource consumption and fulfilling the latency constraints, respectively, where $W_{u} + W_{l} = 1$. A $[W_{u} = 0.5$, $W_{l} = 0.5]$ setting means that both goals are equally important. The weight, $W_{s}$, reflects the priority of fulfilling the latency requirement of slice $s$, and $l_{s}$ is the average latency experienced by the users of slice $s$ during the previous slicing window. Finally, $b_{s}$ represents the bandwidth resources consumed by the users of slice $s$ out of the total bandwidth shared among slices, $B$.

The impact of the function can be tuned by adjusting two parameters, $c_{1}$ and $c_{2}$. Parameter $c_{1}$ determines the steepness of the sigmoid function, thereby specifying the point at which penalties begin to be imposed on the agent's actions. Furthermore, $c_{2}$ denotes the inflection point, which signifies the minimum acceptable delay performance for each slice. In this work, $c_{2}$ reflects the cumulative constraint, $\epsilon$. The values of $c_{1}$ and $c_{2}$ are set for the different slices in accordance with their defined SLAs. Latency is employed as a variable since we prioritize the delay-sensitive VR gaming service and to enhance the interpretability of the results. In order to avoid unsafe actions at the decision time step level, cumulative constraints are reinforced by instantaneous constraints, as explained in the next subsection.

\vspace{2pt}
\subsubsection{Safety Layer}

A feasible action space includes actions leading to instantaneous latency that is less than the defined SLA thresholds. To satisfy the instantaneous latency constraints, we project the inter-slice RA actions generated by the RL agent onto a feasible space. In this work, we introduce a safety layer to the end of the inter-slice RA DRL policy's ANN to perform such a projection. In a given state, the slicing policy decides an RA action and then passes it to the safety layer. Such a layer overrides the action with the nearest feasible alternative if its cost is expected to exceed the defined latency threshold. The safety layer aims to solve \cite{ijcai2021p614}:

\begin{equation}
\label{eq:min_euclidian_distance}
\begin{aligned}
& \min _{a_t^{\prime}} \frac{1}{2}\left\|a_t^{\prime}-a_t\right\|^2 \\
& \text { s.t. } C_i\left(\kappa_t, a_t^{\prime}, \kappa_{t+1}^{\prime}\right) \leq \omega_i
\end{aligned}
\end{equation}

Here, $a_t^{\prime}$ is the new feasible action, $a_t$ is the original infeasible action, $C_i\left(\kappa_t, a_t^{\prime}, \kappa_{t+1}^{\prime}\right)$ is the cost signal under new action $a^{\prime}$ in state $\kappa$ at time $t$, and $\omega_i$ is the predefined instantaneous latency threshold. This is considered a Euclidean distance projection since the action is represented as a vector of bandwidth percentages allocated to each slice, as defined in Section \ref{sec:mapping}.

The new action chosen by the safety layer not only has to satisfy the latency threshold but should also have the smallest distance from the original action among the feasible actions. The latter condition is primarily imposed to ensure that the new action is closely efficient to the original one with respect to resource consumption. A key challenge is that latency is an implicit instantaneous constraint, and its cost function \(C_i(\cdot)\) is unknown. To address this problem, we propose to learn $C_i(\cdot)$ using an SL approach. 

\vspace{2pt}
\subsubsection{Supervised Learning for Cost Prediction}
\label{sec:supervised_model}

We propose to build and employ a supervised regression model to predict the cost of actions given a certain state. This allows the agent to consider the latency resulting from an action on a per-time-step basis when allocating resources to the available slices. The cost signal is represented in our system in terms of the average latency experienced by a slice's users in a slicing window as follows:

\begin{equation}
C_{\Omega_t,s} = \frac{1}{U_s} \sum_{u=1}^{U_s} L_{\Omega_t,s,u}    
\end{equation}

where \( C_{\Omega_t,s} \) is the cost signal for slice \( s \) received after a slicing window \( \Omega_t \), and \( L_{\Omega_t, s, u} \) is the latency experienced by user \( u \) in slice \( s \) during a slicing window \( \Omega_t \).

\begin{figure}
\centering
\includegraphics[width=0.99\linewidth]{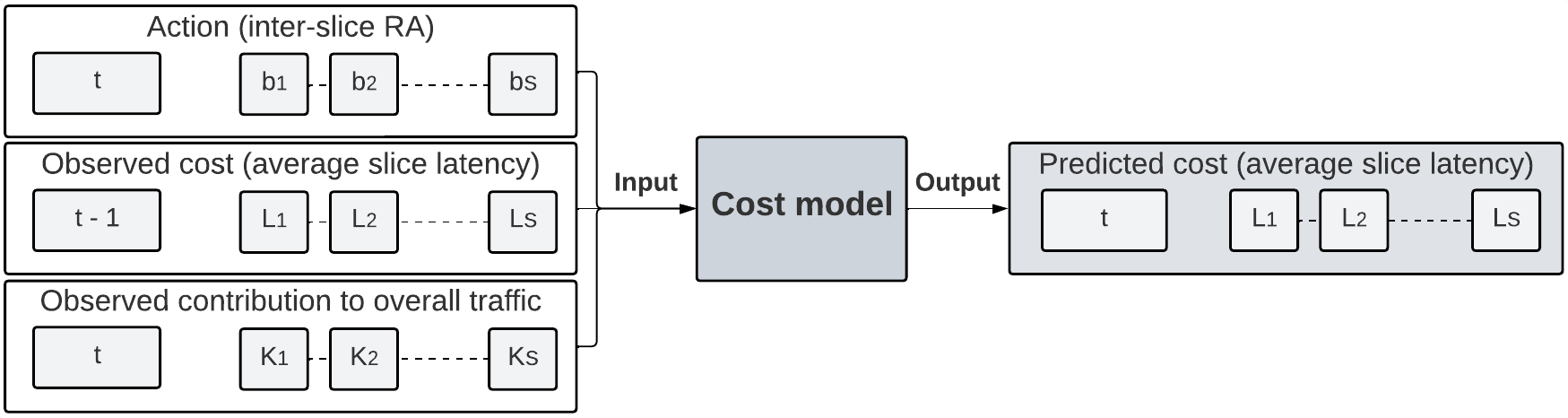}
\caption{Inputs and outputs of the proposed SL cost model.}
\label{fig:proposed_cost_model}
\vspace{-3ex}
\end{figure}

The inputs to the proposed model are the state-action pairs and the cost signal observed from the preceding action, as highlighted in Fig. \ref{fig:proposed_cost_model}. Based on these inputs, the cost model $\mathbb{C}$ predicts the average latency of a slice's users, given that the action $a_t$ suggested by the policy has been taken. This prediction reflects the cost signal expected to be received at the end of the slicing window, $\Omega_{t}$. Consequently, if the predicted cost violates the defined latency thresholds of a slice, the cost model is used to provide a set of feasible actions, \( \mathcal{A}_f \) defined as follows:

\begin{equation}
\label{eq:feasible_action_space}
% A_f = \{a^{\prime} \in \mathcal{A} \mid C(a^{\prime}) < w\}
\mathcal{A}_f = \left\{ a^{\prime} \in \mathcal{A} \ \big| \ \hat{C}_i(\kappa_t, a^{\prime}) \leq \omega_i, \ \forall i \right\}
\end{equation}

The safety layer finally picks the feasible action closest to the original action suggested by the policy as defined in (\ref{eq:min_euclidian_distance}). The proposed approach is summarized in Algorithm \ref{alg:safety_layer}.

\begin{algorithm}
\caption{Proposed Safe DRL-based Inter-slice RA}
\label{alg:safety_layer}
\textbf{Input:} trained cost regression model, $\mathbb{C}$, DRL policy, $\pi$, state, $(\kappa_{1}, . . . ,\kappa_{S})_{t-1}$, slice cost, $(c_{1}, . . . ,c_{S})_{t-1}$, set of possible actions, $\mathcal{A}$, cost threshold, $\omega_{i}$, for each slice \( i \)\\
\textbf{Output:} closest safe action, $a_t^{\prime}$ as defined in Section \ref{sec:safety_deployment_flow}

\begin{algorithmic}[1]

\State \textbf{for} each deployment decision time step \( t_{\text{deploy}} = 1 \) to \( T \) \textbf{do}:
\State \quad  Compute the action suggested by the policy, $a_t = \pi(\kappa_t)$
\State \quad  For each slice \( i \), predict the cost, $\hat{C}_i(\kappa_t, a_t) = \mathbb{C}^*(\kappa_t, a_t)$
\State \quad \textbf{if} there exists any slice \( i \) such that \( \hat{C}_i(\kappa_t, a_t) > \omega_i \) \textbf{then}:
    \State \quad \quad Define the set of feasible safe actions using (\ref{eq:feasible_action_space})
    \State \quad \quad Select the action \( a_t^\prime \in \mathcal{A}_f \) closest to \( a_t \) using (\ref{eq:min_euclidian_distance})
    \State \quad \textbf{else}
    \State \quad \quad Set \( a_t^\prime = a_t \)
\State \quad \textbf{end if}
\State \quad Execute the action \( a_t^\prime \)
\State \quad Observe the reward, actual cost and next state, \( R_t, C_t, \kappa_{t+1} \)
\State \quad Update the agent's policy
\State \textbf{end for}
\end{algorithmic}
\end{algorithm}
\vspace{-2ex}

\begin{table*}
\centering
\caption{Experiment Setup: Simulation Parameters Settings}
\begin{tabular}{|p{1.15in}|p{1.9in}|p{1.665in}|p{1.9in}|}
\hline
\textbf{}   & \textbf{Video}     & \textbf{VoNR}                & \textbf{VR gaming}                                                                                                 \\ \hline

\textbf{Scheduling algorithm}                                                                                                                        & \multicolumn{3}{l|}{Round-robin per 1 ms slot}  \\ \hline
\textbf{Slicing window size}                                                                                                                      & \multicolumn{3}{l|}{PRB allocation among slices every 100 scheduling time slots} \\ 
\hline

\textbf{BS maximum capacity}                                                                                                                    & \multicolumn{3}{l|}{18,750 B per TTI} \\ 
\hline
\textbf{Latency threshold}          & 12 ms  &        20 ms                                               & Pre-training: 10 ms, Testing: 10, 5 ms                                                                    \\
\hline
\textbf{Packet interarrival time}
 & Truncated Pareto (mean = 6 ms, max = 12.5 ms) & Uniform (min = 0 ms, max = 160 ms) & Real VR gaming traces (mean = 11.1 - 18 ms)
   \\ \hline
\textbf{Packet size}
 & Truncated Pareto (mean = 100 B, max = 250 B)      & Constant (40 B)                                                   &  Real VR gaming traces (mean = 43.73 KB - 58.933 KB)
 \\ 
\hline   
\textbf{Number of users }
   & Poisson (mean = 19 - 76)                       & Poisson (mean = 38 - 152)                                                         & Poisson (mean = 1 - 3)\\ \hline
\end{tabular}
\vspace{-3ex}
\label{tab:sim_parameters_exp_safety}
\end{table*}

\subsection{Proposed SLA-Compliant O-RAN Slicing Architecture} 
\label{sec:proposed_approach}

We propose training and deployment workflows that incorporate safe DRL as a core component of radio resource management (RRM) in next-generation O-RANs.

\vspace{1ex}
\subsubsection{Training Workflow}
\label{sec:training_workflow}
The O-RAN architecture provides flexibility in collecting relevant data at different levels and timescales. Hence, we propose to collect data that reflects the state-action-reward-cost experience in the real O-RAN environment. This is to be saved at the non-RT RIC, as highlighted in Fig. \ref{fig:safe_system_diagram}. Such data measures the system's reaction to the inter-slice RA decisions made by the DRL agents. This primarily includes the reward and cost signals received by the DRL agent after taking an RA action given a network condition.

The data is then used to train our proposed cost model, while part of it is used to pre-train DRL agents before their deployment in a real environment. The more data that is collected, the closer the offline training environment resembles a digital twin of the real O-RAN environment. This allows for a more accurate representation of the O-RAN environment and provides historical data to be exploited by the cost model proposed in this paper.

Since we do not have full access to real O-RAN interactions, we follow a guided DR approach to imitate the proposed data collection process at the non-RT RIC. Guided DR conserves computational resources by avoiding training in unrealistic conditions. Instead of randomly performing a grid search over various RAs at different traffic levels, we use both realistic and model-based datasets to filter the traffic levels in such a DR process. This includes real VR gaming traces \cite{9685808} and common mathematical traffic models of voice over new radio (VoNR) and video services. Various discretized levels of inter-slice RA actions are then explored to compute the system response under varying network conditions. Consequently, the collected data is used to pre-train the DRL agents and the proposed cost model. The different scenarios addressed in this paper are detailed in Section \ref{sec:simulation_setup_safety}.

\subsubsection{Deployment Workflow}
\label{sec:safety_deployment_flow}

Algorithm \ref{alg:safety_layer} and Fig. \ref{fig:safe_system_diagram} summarize the control sequence of the proposed approach. At each decision time step $t$, the learner agent observes the system state via O-RAN’s E2 interface in addition to the SLA configurations of each slice as set by the MNO through the A1 interface. The agent then selects an inter-slice RA configuration based on the state representation vector reflecting the contribution of each slice to the overall traffic as defined in Section \ref{sec:mapping}.

This step is followed by consulting the trained cost model via O-RAN’s A1 interface. The model takes the state-action pair in addition to the previous cost as input and predicts the cost of such an action. If the expected cost exceeds the latency threshold defined by the MNO, the safety layer masks actions, allowing only those expected to have a cost below the defined threshold. It then chooses the one with the smallest Euclidean distance to the original action.

The action satisfying such conditions is then communicated with the virtualized O-RAN environment via the E2 interface. Hence, the system operates using such an inter-slice RA configuration throughout the slicing window duration until the next decision time step $t+1$. The intra-slice RA, i.e., scheduling, is performed among the users of each slice during this time. At the end of the slicing window, the agent collects the reward and cost signals reflecting the system's performance with regard to its goals and constraints. The objective of this control loop is to allocate the PRBs efficiently while abiding by the SLAs defined by the MNO with high probability.

\vspace{-1ex}

\section{Experiment Setup and Numerical Results}
\label{sec:results}

\subsection{Experiment Setup}
\label{sec:simulation_setup_safety}

In this performance study, we carry out guided DR as described in Section \ref{sec:training_workflow}. This simulates various situations a DRL agent experiences when newly deployed to a network or when there is a change in the network conditions. We base such a DR on models defined in Table \ref{tab:sim_parameters_exp_safety}. The VR gaming traces used in this paper are based on two heterogeneous games played under different data rate limits, as explained in \cite{9685808}. We first pre-train all the DRL agents using the train split of the three services traffic. We then use the pre-trained policies to carry out inter-slice RA when experiencing demands based on the test split of the generated traffic. We evaluate the system's performance based on these test situations, as it is more challenging for a DRL agent to adapt when changes happen to the network. We define traffic levels based on the number of users in each slice and the average VR gaming traffic demand per user per slicing window.

We also experiment with different VR gaming latency thresholds to evaluate the performance of the proposed method and baselines under changing constraints. The simulation settings, including ranges of user counts and the slices' latency thresholds used in this study, are defined in Table \ref{tab:sim_parameters_exp_safety}. With the defined training-deployment scenarios, the test cases can be categorized into the following four high-level classes:
\begin{enumerate}
    \item A policy is tested in an environment with traffic demands and latency threshold similar to the environment in which it was pre-trained.
    \item A policy is tested in an environment with traffic demands similar to the environment in which it was pre-trained. However, it has a more restricted latency threshold. 
    \item A policy is tested in an environment with traffic demands different from the environment in which it was pre-trained. However, it has a similar latency threshold. 
    \item A policy is tested in an environment with different traffic demands and a more restricted latency threshold than the environment in which it was pre-trained. 
\end{enumerate}

\paragraph{DRL algorithm}

Our approach is algorithm-agnostic. We demonstrate it by modifying the advantage actor-critic (A2C) DRL algorithm, commonly used in the slicing literature. We refer to such a modified version as SafeSlice. Table \ref{tab:drl_parameters_safety} highlights the main DRL settings used in this performance study.

\paragraph{Cost Model} We use the extreme gradient boosting (XGBoost) algorithm to build a model for slice cost prediction. XGBoost is an implementation of gradient boosting optimized for performance, scalability, and speed in deployment, which is our main concern. We use XGBoost implementation from the Distributed (Deep) Machine Learning Community (DMLC)'s GitHub repository\footnote{\href{https://github.com/dmlc/xgboost}{Available at https://github.com/dmlc/xgboost}}. In this study, we train the proposed XGBoost cost model by combining DMLC's XGBoost implementation and RAPIDS cuML\footnote{\href{https://github.com/rapidsai/cuml}{Available at https://github.com/rapidsai/cuml}}, which is a suite of GPU-accelerated ML algorithms. We employ an automated ML (AutoML) tool called TPOT \cite{le2020scaling} to realize such a combination and train several XGBoost models using multiple hyperparameter settings. We configure TPOT to optimize such settings to minimize the regression error of the trained models. We then follow a 5-fold cross-validation approach and calculate the performance metrics listed in Table \ref{tab:xgboost} using the model with the best hyperparameter settings among the ones tested.

\begin{table}
\centering
\caption{Experiment Setup: RAN slicing DRL settings.}
\begin{tabular}{p{1.18in}p{2.1in}}
\hline
\textbf{Setting}                                                                                                                      & \textbf{Value}                                                                                                            \\ 
\hline
\textbf{Reward function}                                                                                                                       & \textbf{A2C and SafeSlice}: The multi-objective reward function defined in (\ref{eqn:safetyrewardequation}), \textbf{SAC-L}: Only the part minimizing resource consumption in (\ref{eqn:safetyrewardequation})                                               \\ 
\textbf{Cost definition}                                                                      & Average latency experienced by the VR gaming service users                                                                                                         \\ 
\textbf{Reward function weights}                                                                                                                      & $W_{u}=0.5$, $W_{l}=0.5$                                                                                                                                                             \\ 
\textbf{DRL algorithm}                                                                                                                       & A2C, Proposed SafeSlice, SAC-L                                                                                                                        \\ 
\textbf{Learning steps per run}                                                                                                                       & \textbf{Training}: 20,000 - 40,000, \textbf{Testing}: 10,000                                                                                                                            \\ 
\textbf{Discount factor}                                                                                                                       & 0.7 - 0.95                                                                                                                                              \\                                  

\textbf{Learning rate}                                                                                                                     & 0.0005 - 0.01                                                                                                                                 \\ 

\textbf{Batch size}                                                                                                                   & 200                                                                                                                                                               \\ 
\hline

\end{tabular}
\vspace{-2ex}
\label{tab:drl_parameters_safety}
\end{table}

\paragraph{Baselines}

1) \textit{Unconstrained A2C:} we use A2C as the unconstrained RL baseline. For a fair comparison, we use the same reward function (\ref{eqn:safetyrewardequation}) in both the proposed SafeSlice approach and the unconstrained A2C. 2) \textit{SafeSlice with Perfect Predictor:} We also evaluate the performance of our proposed approach when using a perfect cost model to predict the instantaneous cost and enforce alternative safe actions to avoid SLA violations. 3) \textit{SAC-Lagrangian:} We finally adopt SAC-Lagrangian \cite{pmlr-v155-ha21c} as a baseline and refer to it as SAC-L. SAC-L-based approaches have recently been proposed in several slicing studies, as reviewed in Section \ref{sec:related_safety}. SAC-L combines SAC with Lagrangian methods to address instantaneous constraints. As shown in Table \ref{tab:drl_parameters_safety}, the SAC-L reward function is modified to exclude the latency constraints, which are instead captured in a separate cost function following the constrained RL framework. We also use the same value for both cumulative and instantaneous latency thresholds, $\epsilon$ and $\omega$. These settings facilitate comparing methods that embed constraints directly in the reward function with those that follow the constrained RL approach over both short and long horizons.

\begin{table}
\centering
\caption{Trained regression model for cost prediction. \label{tab:xgboost}}
\begin{tabular}{p{1.15in}p{2.1in}}
\hline
\multicolumn{2}{c}{\textbf{Features}} \\ \hline
\textbf{Input}       & Observed state, inters-slice RA action, and observed cost        \\ 
\textbf{Output}      & Expected cost of the original action suggested by the policy                         \\ 
\textbf{Estimator}   & XGBRegressor             \\ 
\hline
\multicolumn{2}{c}{\textbf{Hyperparameters}} \\ 
\hline
\textbf{Alpha}       & 10                       \\ 
\textbf{Learning rate} & 0.5                   \\ 
\textbf{Max depth}  & 9                        \\ 
\textbf{Minimum child weight} & 14                \\ 
\textbf{Number of estimators} & 100                    \\ 
\textbf{Objective}   & Squared error         \\ 
\textbf{Subsample}   & 1.0                      \\ 
\hline

\multicolumn{2}{c}{\textbf{Performance}} \\ 
\hline
\textbf{RMSE on test data}  & 4.42 (average latency $\in [0, 50]$) \\ 
\textbf{R-Squared score}    & 0.94              \\ 
\hline
\end{tabular}
\vspace{-2ex}
\end{table}

\vspace{-0.6ex}

\subsection{Numerical Results}

\vspace{-1ex}

\paragraph{Cumulative Latency Performance}

\begin{figure} 
  \begin{subfigure}[b]{0.49\linewidth}
    \centering
\includegraphics[width=1\linewidth]{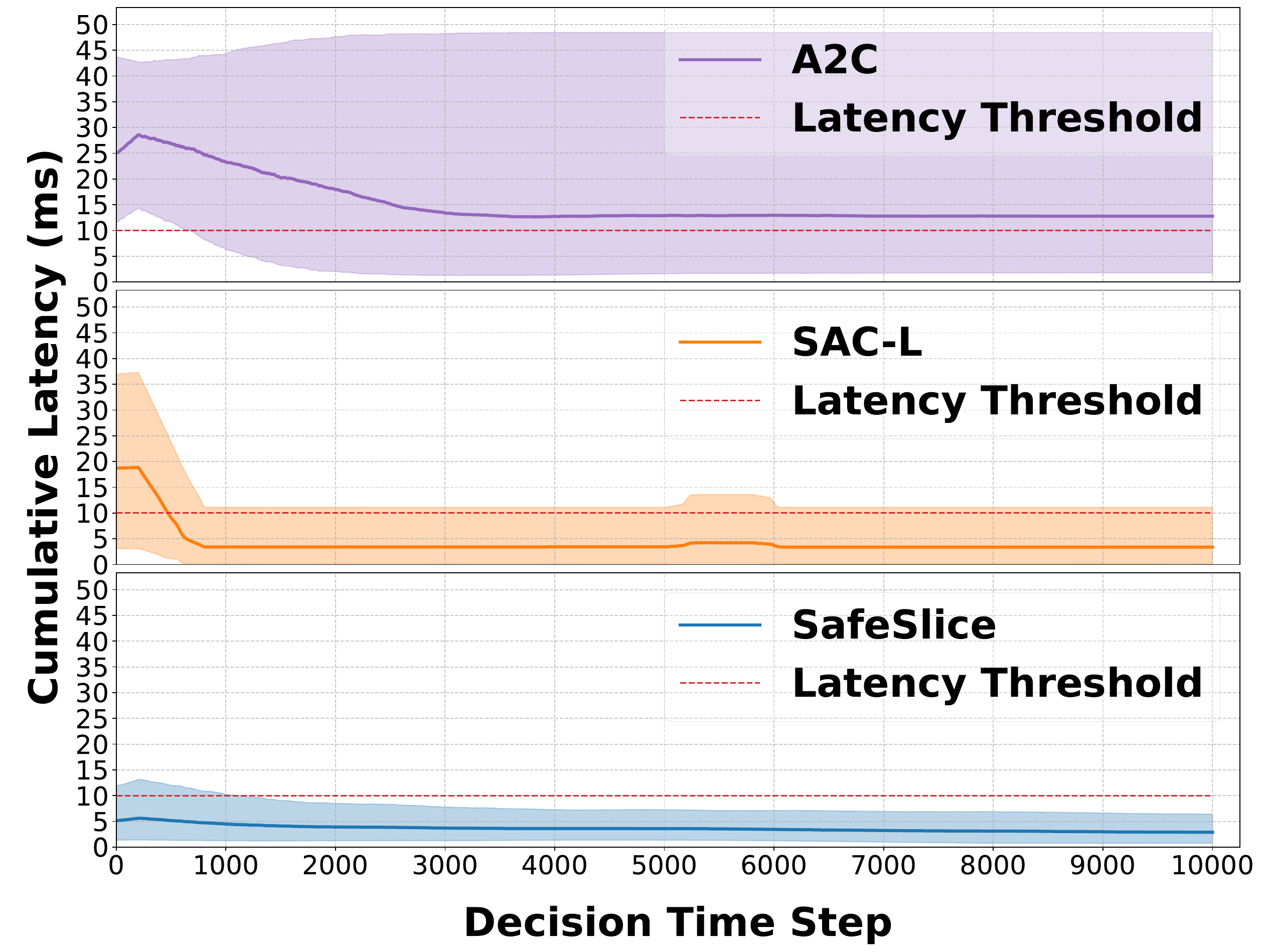} 
\vspace{-4ex}
    \caption{\scriptsize Same traffic and latency threshold.}
    \label{fig:cost_same_same} 
  \end{subfigure}
  \begin{subfigure}[b]{0.49\linewidth}
    \centering
    \includegraphics[width=1\linewidth]{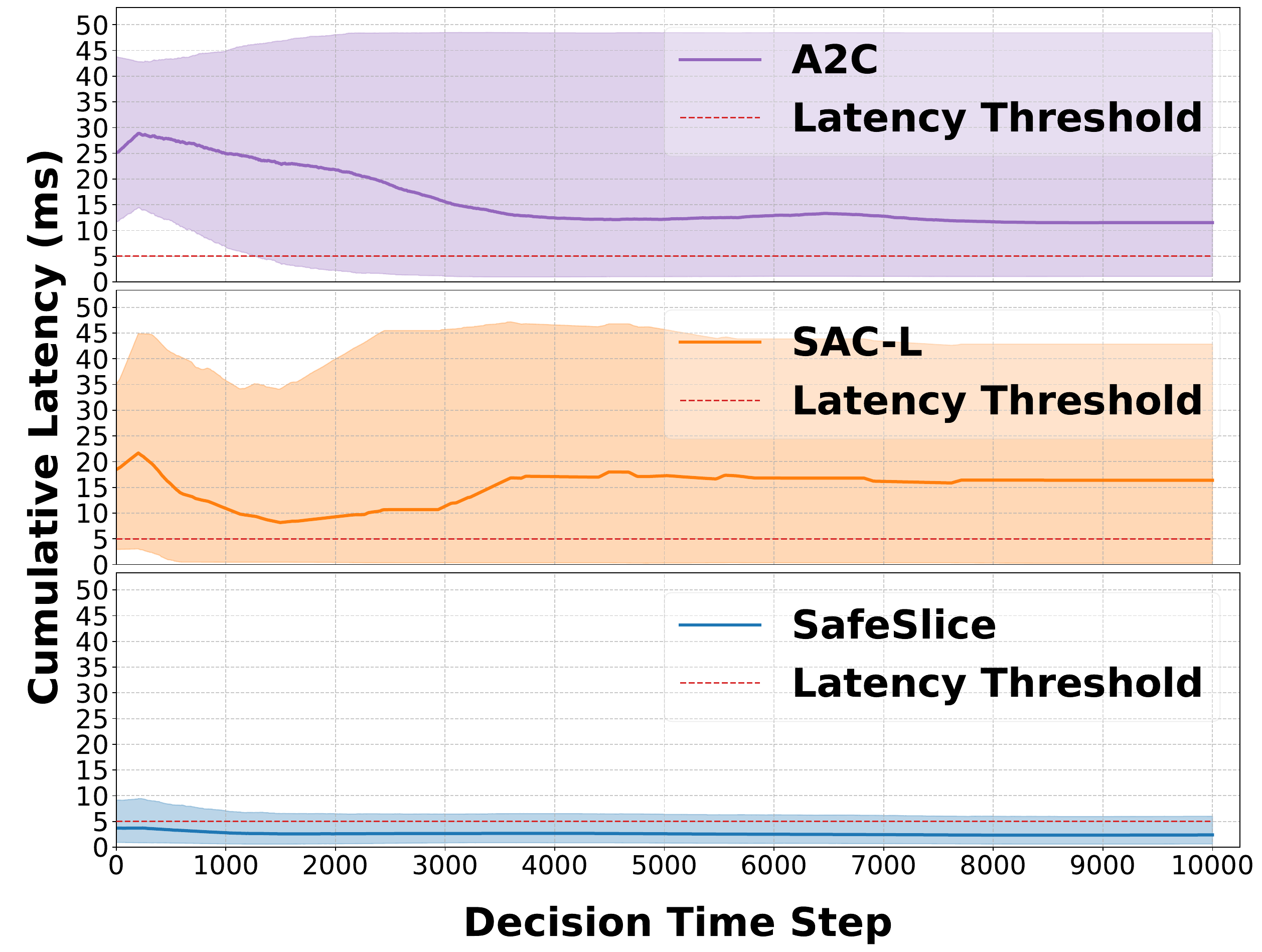} 
\vspace{-4ex}
    \caption{\scriptsize Same traffic, different latency threshold.} 
    \label{fig:cost_same_diff} 
  \end{subfigure} 
  \begin{subfigure}[b]{0.49\linewidth}
    \centering
    \includegraphics[width=1\linewidth]{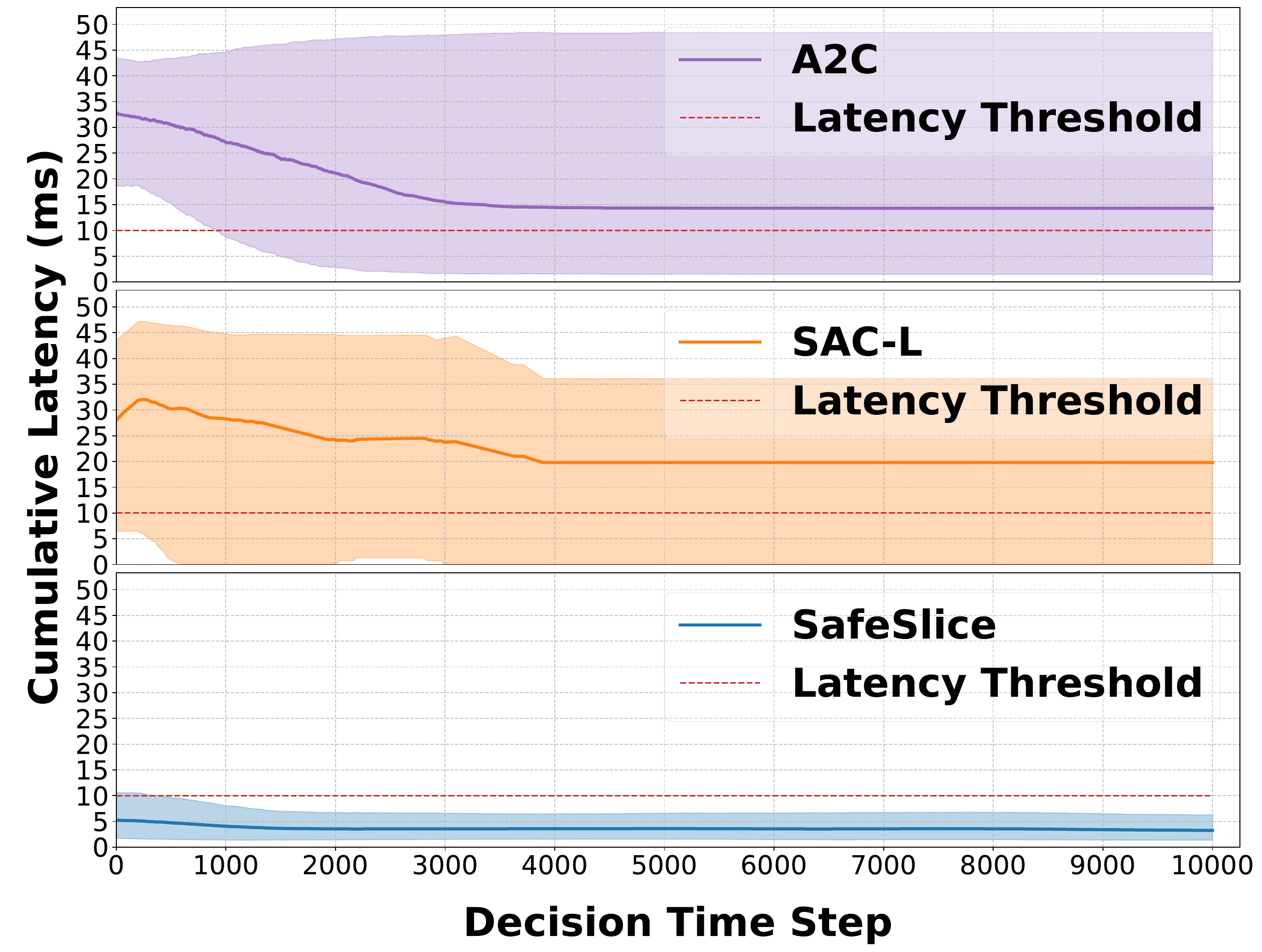} 
\vspace{-4ex}
    \caption{\scriptsize Different traffic, same latency threshold.} 
    \label{fig:cost_diff_same} 
  \end{subfigure}
  \begin{subfigure}[b]{0.49\linewidth}
    \centering
    \includegraphics[width=1\linewidth]{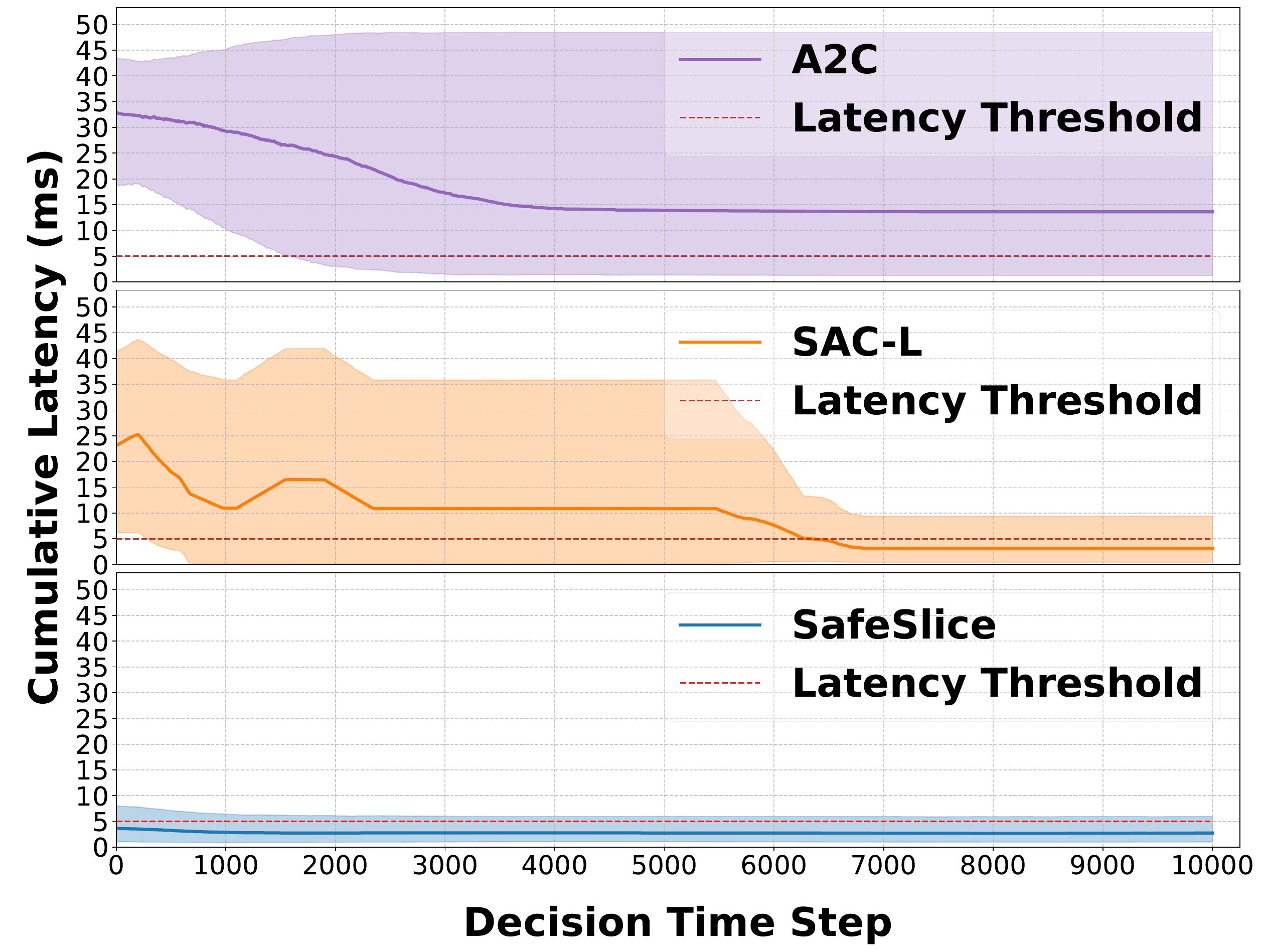} 
\vspace{-4ex}
    \caption{\scriptsize Different traffic and latency threshold.} 
\label{fig:cost_diff_diff} 
  \end{subfigure}
      \vspace{-2ex}
\vspace{-1.5ex}
 \setlength{\belowcaptionskip}{-15pt} 
  \caption{Cumulative latency under the four traffic test categories.}
  \label{fig:overall_cost_performance}
\end{figure}

We first examine the long-term average latency of the VR gaming service users in Fig. \ref{fig:overall_cost_performance}. SafeSlice abides by the long-term latency constraints. It instantaneously overrides actions that may lead to violating the defined latency threshold. This happens at the cost of more resource consumption than unconstrained A2C, as discussed later. SAC-L performs poorly when the deployment environment changes, especially when only one variable is modified. For instance, its performance declines when tested with the same latency threshold as the training environment but with different traffic demands. Since the threshold is the same, it uses its trained policy's ANN as a function approximator for the optimal actions. However, the previously unseen significant changes in traffic demands render the ANN's predictions inaccurate. Furthermore, the unconstrained A2C fails to comply with the long-term constraints in all scenario categories. This shows that embedding the constraints into a multi-objective reward function is insufficient.

\begin{figure} 
  \begin{subfigure}[b]{0.5\linewidth}
    \centering
\includegraphics[width=1\linewidth]{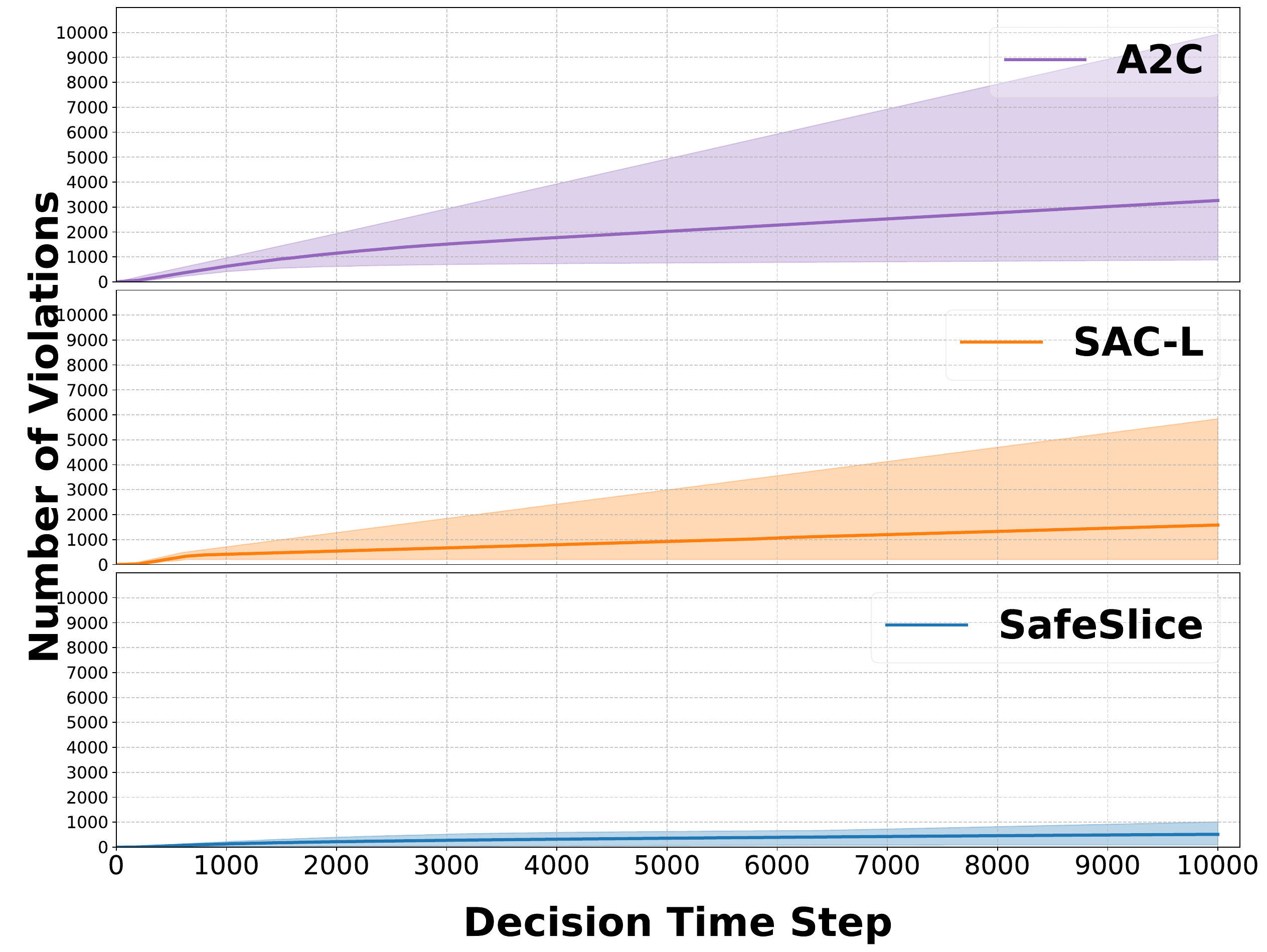} 
\vspace{-4ex}
    \caption{\scriptsize Same traffic and latency threshold.}
    \label{fig:cum_same_same} 
  \end{subfigure}
  \begin{subfigure}[b]{0.5\linewidth}
    \centering
    \includegraphics[width=1\linewidth]{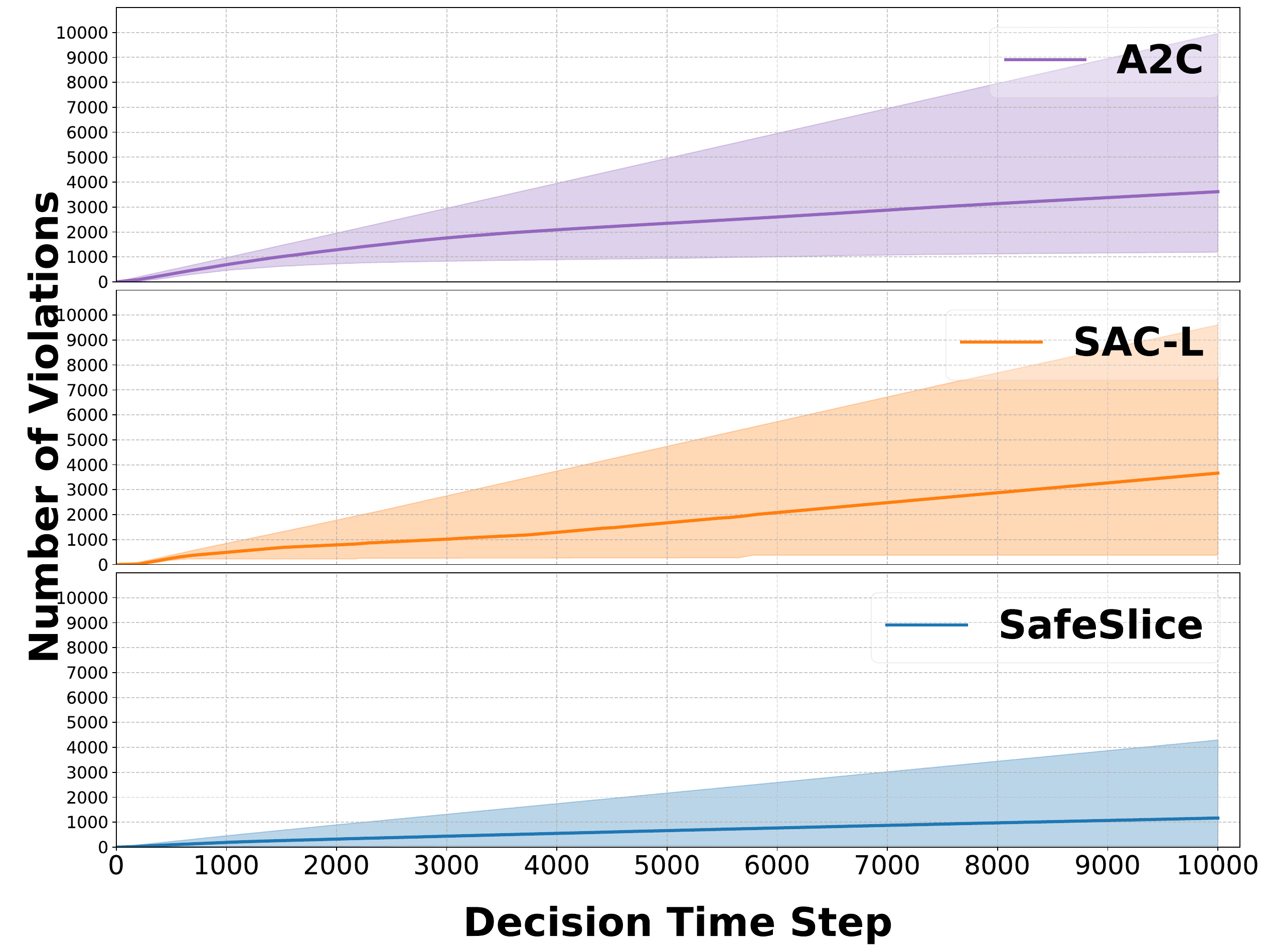} 
\vspace{-4ex}
    \caption{\scriptsize Same traffic, different latency threshold.} 
    \label{fig:cum_same_diff} 
  \end{subfigure} 
  \begin{subfigure}[b]{0.5\linewidth}
    \centering
    \includegraphics[width=1\linewidth]{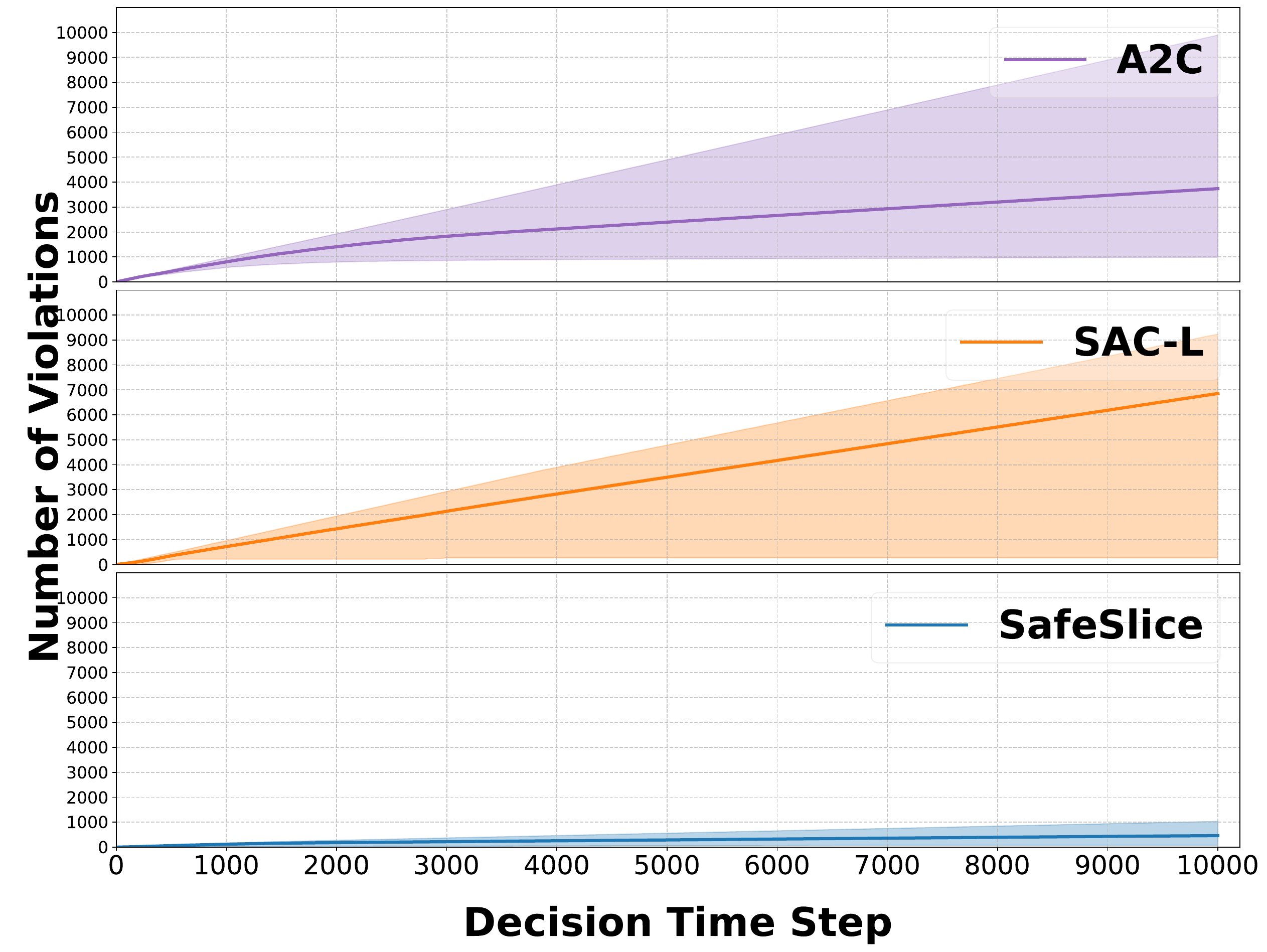} 
\vspace{-4ex}
    \caption{\scriptsize Different traffic, same latency threshold.} 
    \label{fig:cum_diff_same} 
  \end{subfigure}
  \begin{subfigure}[b]{0.5\linewidth}
    \centering
    \includegraphics[width=1\linewidth]{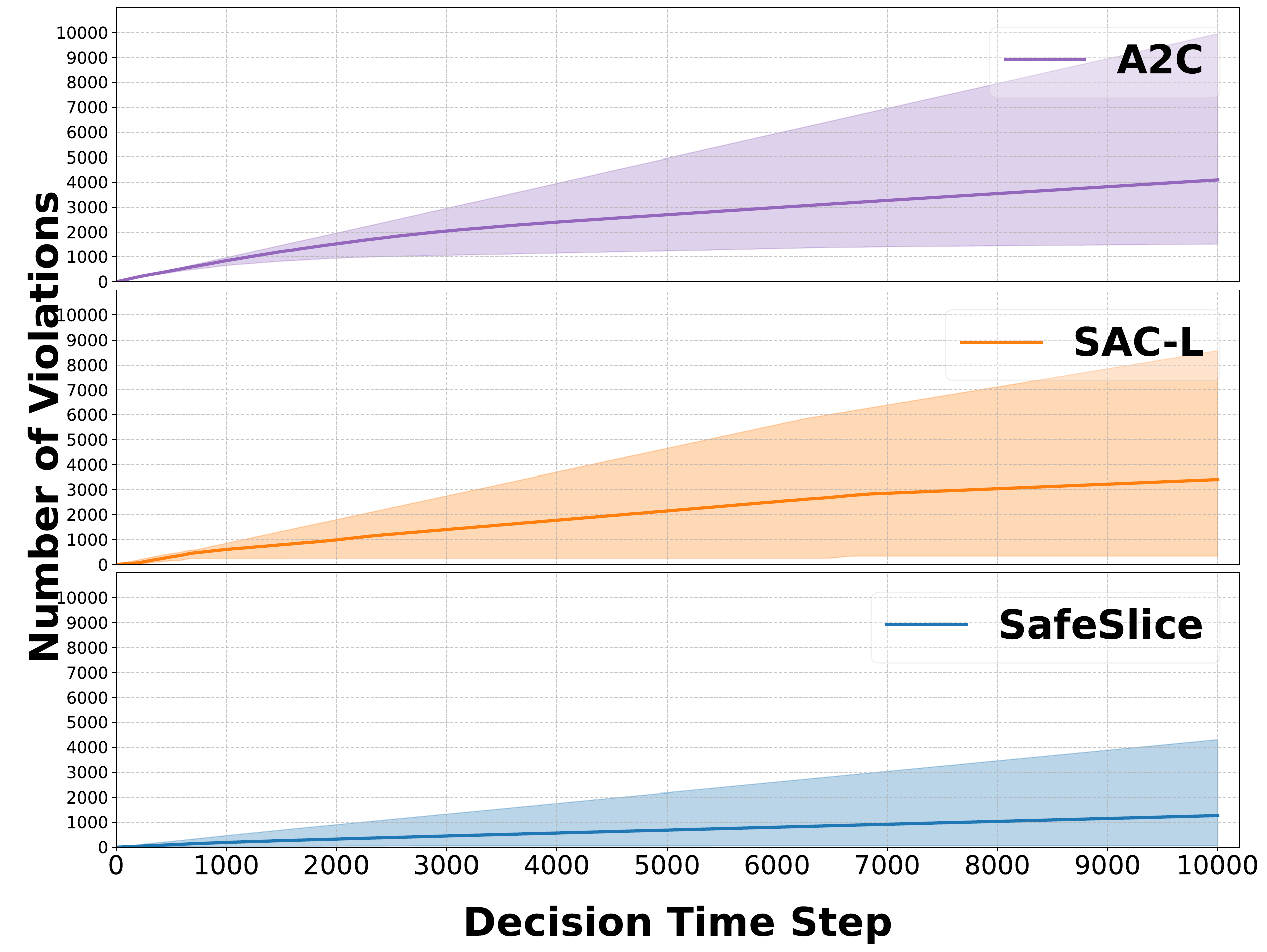} 
\vspace{-4ex}
    \caption{\scriptsize Different traffic and latency threshold.} 
\label{fig:cum_diff_diff} 
  \end{subfigure}
      \vspace{-2ex}
\vspace{-1.5ex}
 \setlength{\belowcaptionskip}{-10pt} 
\caption{Instantaneous violations accumulated over time.}
  \label{fig:overall_cum_violations_performance}
\end{figure}

\paragraph{Violations Performance}
\label{sec:error_effect}

We then investigate the number of instantaneous violations accumulated over time in Fig. \ref{fig:overall_cum_violations_performance}. SafeSlice performs significantly better in all four categories of test scenarios. It manages to keep a low violation rate even when experiencing changes in traffic and latency threshold configurations. However, it performs relatively worse when the threshold configuration is changed than in the other situations. This could be due to the cost model's higher sensitivity to error with a lower latency threshold, especially given that its RMSE is 4.42. It may also be unable to capture the dynamics of such a scenario compared to the others, given the training data. SAC-L performs well only when similar conditions are experienced. It struggles to enforce instantaneous constraints when experiencing new scenarios. It cannot adapt quickly to online changes in traffic load and latency thresholds. The unconstrained A2C performs almost the same in all the categories, as it only deals with constraints as part of the reward function.

\paragraph{Resource Consumption Performance}

\vspace{-0.4ex}

\begin{figure} 
  \begin{subfigure}[b]{0.5\linewidth}
    \centering
\includegraphics[width=1\linewidth]{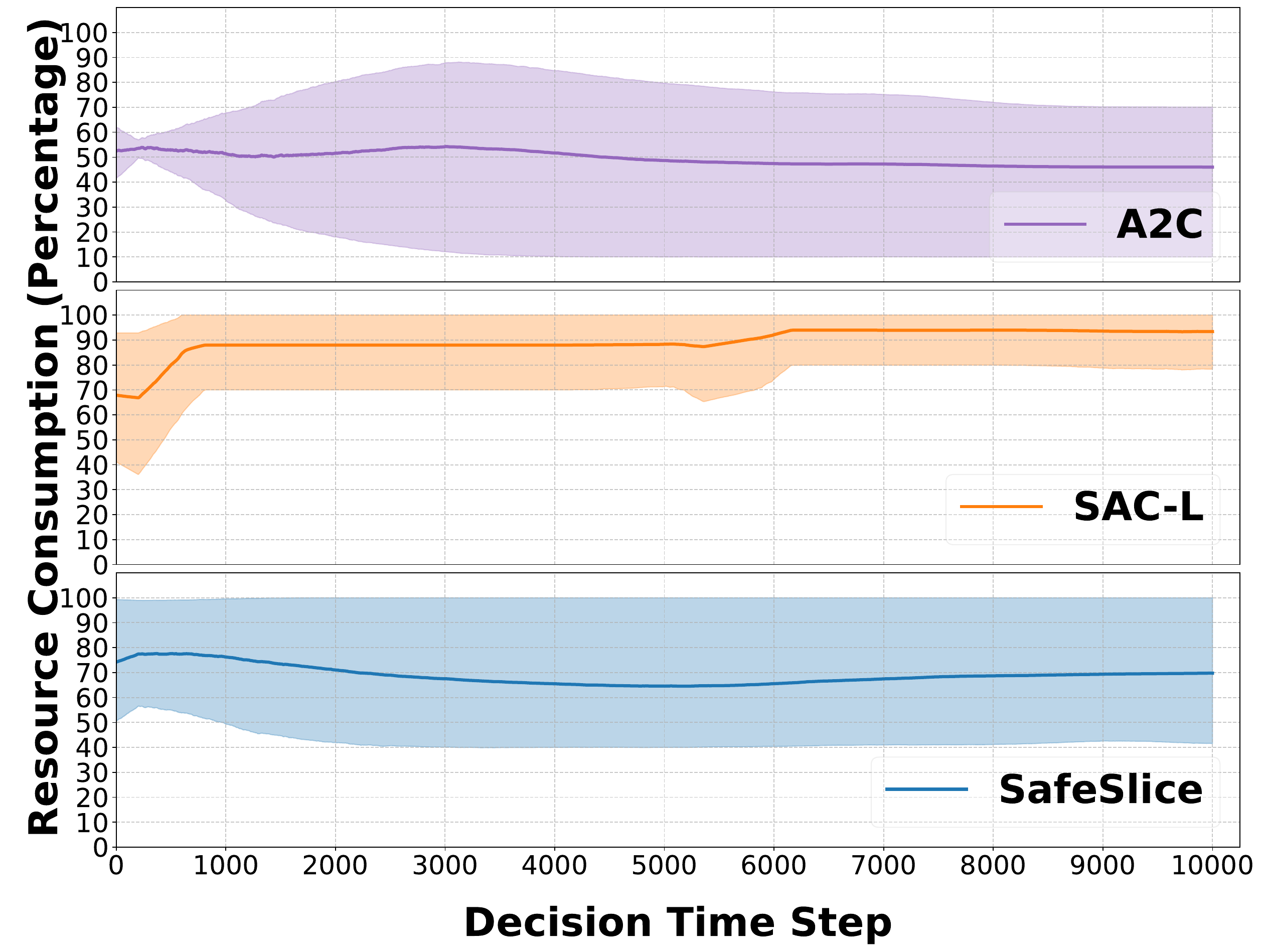} 
\vspace{-4ex}
    \caption{\scriptsize Same traffic and latency threshold.}
    \label{fig:allocation_same_same} 
  \end{subfigure}
  \begin{subfigure}[b]{0.5\linewidth}
    \centering
    \includegraphics[width=1\linewidth]{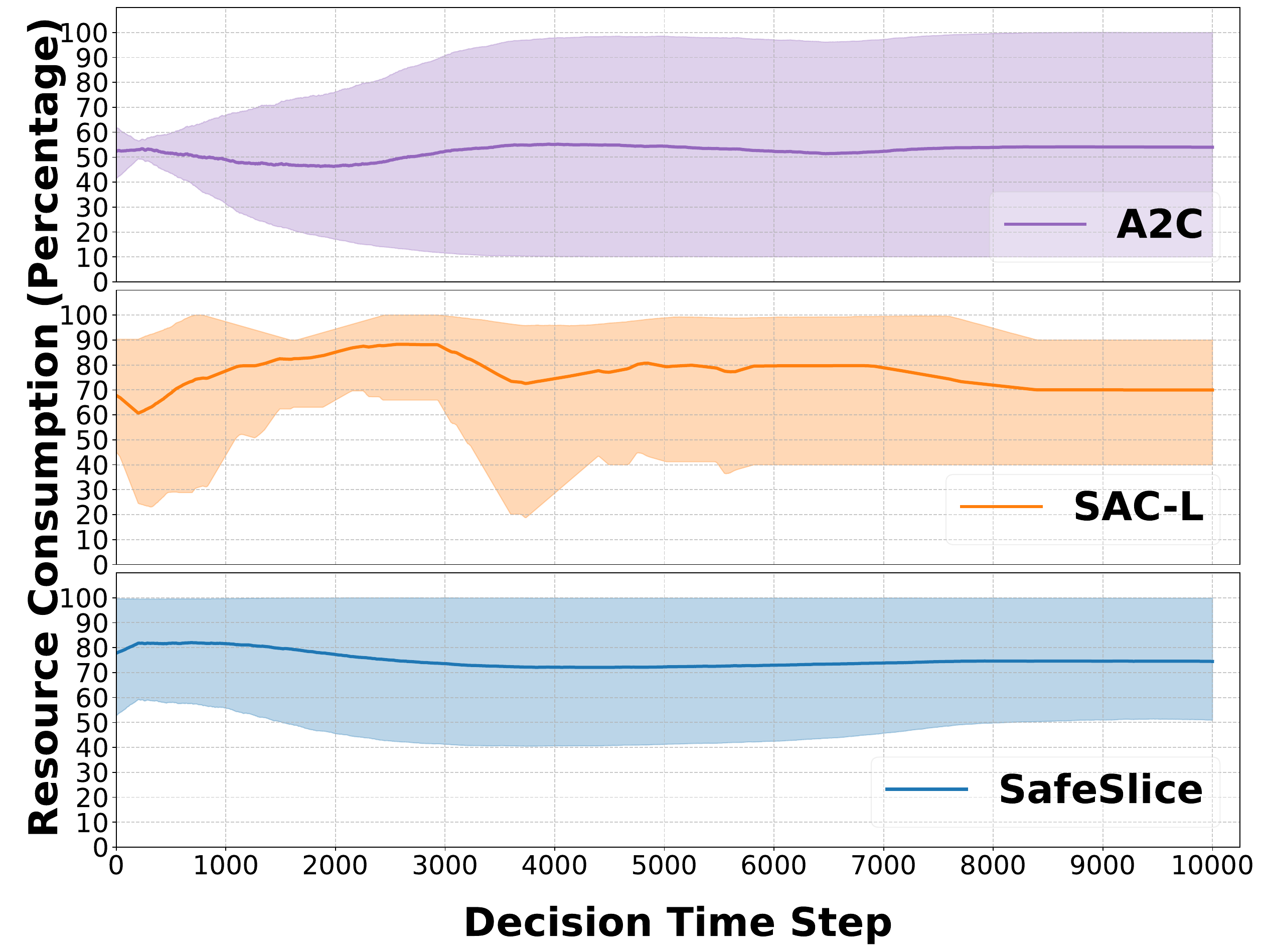} 
\vspace{-4ex}
    \caption{\scriptsize Same traffic, different latency threshold.} 
    \label{fig:allocation_same_diff} 
  \end{subfigure} 
  \begin{subfigure}[b]{0.5\linewidth}
    \centering
    \includegraphics[width=1\linewidth]{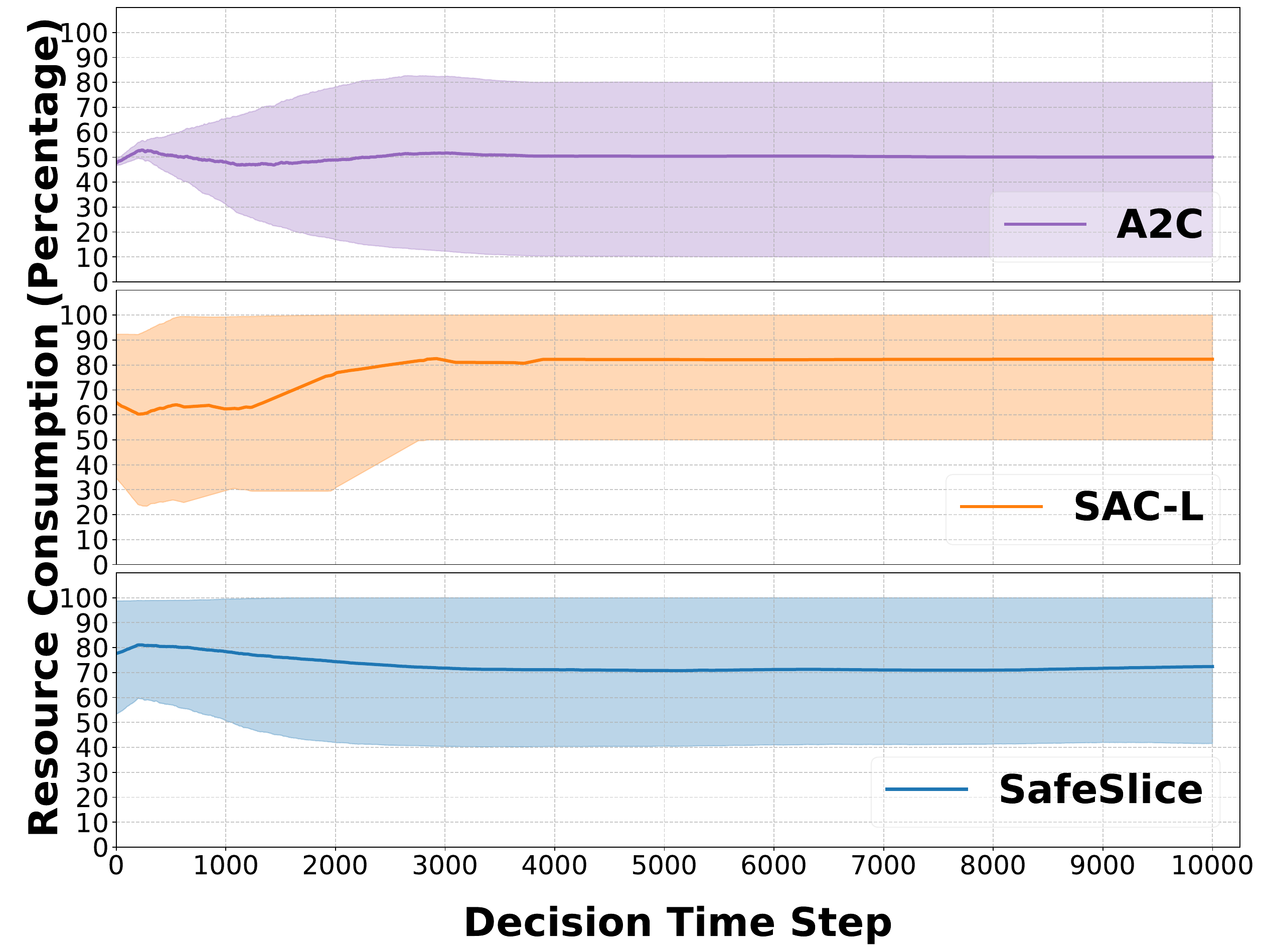} 
\vspace{-4ex}
    \caption{\scriptsize Different traffic, same latency threshold.} 
    \label{fig:allocation_diff_same} 
  \end{subfigure}
  \begin{subfigure}[b]{0.5\linewidth}
    \centering
    \includegraphics[width=1\linewidth]{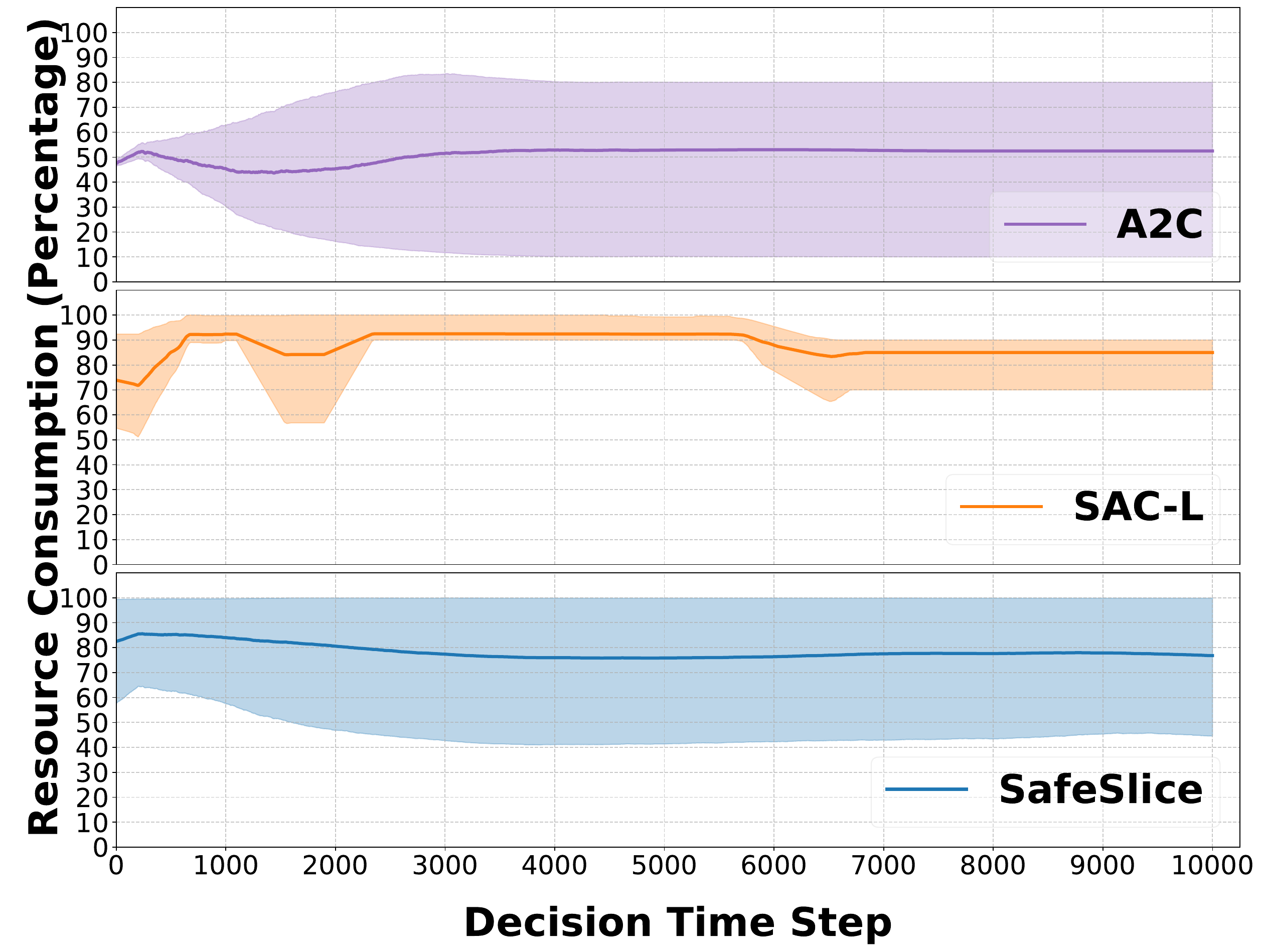} 
\vspace{-4ex}
    \caption{\scriptsize Different traffic and latency threshold.} 
\label{fig:allocation_diff_diff} 
  \end{subfigure}
      \vspace{-2ex}
\vspace{-1.5ex}
 \setlength{\belowcaptionskip}{-15pt} 
  \caption{Resource consumption under the four test categories.}
  \label{fig:overall_allocation_performance}
\end{figure}

We finally examine resource consumption over time in Fig. \ref{fig:overall_allocation_performance}. SafeSlice is slightly better than the SAC-L algorithm in the four test categories. SAC-L exhausts resources unnecessarily, and unconstrained A2C is the best, but this comes at the considerable cost of failed constraint enforcement. Overall, our method achieves reductions of up to 83.23\% in average cumulative cost, 93.24\% in instantaneous latency violations, and 22.13\% in resource consumption compared to the baselines. The results also confirmed that SafeSlice has a very close performance to that when a perfect cost model is utilized, but the figure was omitted due to space limitation.

\section{Conclusion and Future Work}
\label{sec:conclusion}

This paper introduces SafeSlice, a DRL approach that optimizes resource consumption while adhering to cumulative and instantaneous latency constraints. This is achieved by designing a risk-sensitive reward function and building a regression model that guides the DRL agent in selecting a safe action. SafeSlice reduces the average cumulative latency and instantaneous latency violations with acceptable resource consumption. The study indicates that relying solely on a multi-objective reward function to reflect the constraints is insufficient. Although SAC-L is frequently used in the slicing literature, the results show that it fails to adapt its policy on the fly in dynamic O-RAN environments. Future research should integrate multiple methods for a comprehensive, trustworthy DRL-based solution. For instance, transfer learning and constrained DRL can be combined to enhance generalization while maintaining safety guarantees.

\bibliographystyle{IEEEtran}
\bibliography{references.bib}

\begin{thebibliography}{10}
\providecommand{\url}[1]{#1}
\csname url@samestyle\endcsname
\providecommand{\newblock}{\relax}
\providecommand{\bibinfo}[2]{#2}
\providecommand{\BIBentrySTDinterwordspacing}{\spaceskip=0pt\relax}
\providecommand{\BIBentryALTinterwordstretchfactor}{4}
\providecommand{\BIBentryALTinterwordspacing}{\spaceskip=\fontdimen2\font plus
\BIBentryALTinterwordstretchfactor\fontdimen3\font minus \fontdimen4\font\relax}
\providecommand{\BIBforeignlanguage}[2]{{%
\expandafter\ifx\csname l@#1\endcsname\relax
\typeout{** WARNING: IEEEtran.bst: No hyphenation pattern has been}%
\typeout{** loaded for the language `#1'. Using the pattern for}%
\typeout{** the default language instead.}%
\else
\language=\csname l@#1\endcsname
\fi
#2}}
\providecommand{\BIBdecl}{\relax}
\BIBdecl

\bibitem{10054381}
C.-X. Wang, X.~You, X.~Gao, X.~Zhu, Z.~Li, C.~Zhang, H.~Wang, Y.~Huang, Y.~Chen, H.~Haas, J.~S. Thompson, E.~G. Larsson, M.~D. Renzo, W.~Tong, P.~Zhu, X.~Shen, H.~V. Poor, and L.~Hanzo, ``{On the Road to 6G: Visions, Requirements, Key Technologies, and Testbeds},'' \emph{IEEE Communications Surveys \& Tutorials}, vol.~25, no.~2, pp. 905--974, 2023.

\bibitem{9903386}
A.~M. Nagib, H.~Abou-zeid, and H.~S. Hassanein, ``{Toward Safe and Accelerated Deep Reinforcement Learning for Next-Generation Wireless Networks},'' \emph{IEEE Network}, vol.~37, no.~2, pp. 182--189, 2023.

\bibitem{o-ran-specification-slicing-architecture}
{O-RAN Working Group 1}, ``{O-RAN Slicing Architecture 12.0},'' document O-RAN.WG1.Slicing-Architecture-R003-v12.00, O-RAN Alliance, Alfter, Germany, 2024.

\bibitem{9627832}
L.~Bonati, S.~D'Oro, M.~Polese, S.~Basagni, and T.~Melodia, ``{Intelligence and Learning in O-RAN for Data-Driven NextG Cellular Networks},'' \emph{IEEE Communications Magazine}, vol.~59, no.~10, pp. 21--27, 2021.

\bibitem{ijcai2021p614}
Y.~Liu, A.~Halev, and X.~Liu, ``{Policy Learning with Constraints in Model-free Reinforcement Learning: A Survey},'' in \emph{Proceedings of the Thirtieth International Joint Conference on Artificial Intelligence (IJCAI)}, 2021, pp. 4508--4515.

\bibitem{10.1145/3485983.3494850}
Q.~Liu, N.~Choi, and T.~Han, ``{OnSlicing: online end-to-end network slicing with reinforcement learning},'' in \emph{Proceedings of the 17th International Conference on Emerging Networking EXperiments and Technologies}, 2021, p. 141–153.

\bibitem{10329913}
M.~Zangooei, M.~Golkarifard, M.~Rouili, N.~Saha, and R.~Boutaba, ``{Flexible RAN Slicing in Open RAN With Constrained Multi-Agent Reinforcement Learning},'' \emph{IEEE Journal on Selected Areas in Communications}, vol.~42, no.~2, pp. 280--294, 2024.

\bibitem{10154267}
M.~Sulaiman, M.~Ahmadi, M.~A. Salahuddin, R.~Boutaba, and A.~Saleh, ``{Generalizable Resource Scaling of 5G Slices using Constrained Reinforcement Learning},'' in \emph{IEEE/IFIP Network Operations and Management Symposium (NOMS)}, 2023, pp. 1--9.

\bibitem{9333595}
Y.~Xu, Z.~Zhao, P.~Cheng, Z.~Chen, M.~Ding, B.~Vucetic, and Y.~Li, ``{Constrained Reinforcement Learning for Resource Allocation in Network Slicing},'' \emph{IEEE Communications Letters}, vol.~25, no.~5, pp. 1554--1558, 2021.

\bibitem{9671840}
Y.~Liu, J.~Ding, Z.-L. Zhang, and X.~Liu, ``{CLARA: A Constrained Reinforcement Learning Based Resource Allocation Framework for Network Slicing},'' in \emph{IEEE International Conference on Big Data}, 2021, pp. 1427--1437.

\bibitem{10437780}
A.~M. Nagib, H.~Abou-zeid, and H.~S. Hassanein, ``{How Does Forecasting Affect the Convergence of DRL Techniques in O-RAN Slicing?}'' in \emph{IEEE Global Communications Conference (GLOBECOM)}, 2023, pp. 2644--2649.

\bibitem{9685808}
S.~Zhao, H.~Abou-zeid, R.~Atawia, Y.~S.~K. Manjunath, A.~B. Sediq, and X.-P. Zhang, ``{Virtual Reality Gaming on the Cloud: A Reality Check},'' in \emph{IEEE Global Communications Conference (GLOBECOM)}, 2021, pp. 1--6.

\bibitem{le2020scaling}
T.~T. Le, W.~Fu, and J.~H. Moore, ``{Scaling Tree-Based Automated Machine Learning to Biomedical Big Data with a Feature Set Selector},'' \emph{Bioinformatics}, vol.~36, no.~1, pp. 250--256, 2019.

\bibitem{pmlr-v155-ha21c}
S.~Ha, P.~Xu, Z.~Tan, S.~Levine, and J.~Tan, ``{Learning to Walk in the Real World with Minimal Human Effort},'' in \emph{Proceedings of the 2020 Conference on Robot Learning}, vol. 155, 2021, pp. 1110--1120.

\end{thebibliography}

\end{document}